\documentclass[conference, 10pt]{IEEEtran}
\IEEEoverridecommandlockouts
\usepackage{cite}
\usepackage{amsmath,amssymb,amsfonts}
\usepackage{algorithmic}
\usepackage{graphicx}
\usepackage{textcomp}
\usepackage{xcolor}
\usepackage{hyperref}
\usepackage{float}
\usepackage{caption}

\def\BibTeX{{\rm B\kern-.05em{\sc i\kern-.025em b}\kern-.08em
    T\kern-.1667em\lower.7ex\hbox{E}\kern-.125emX}}
\begin{document}

\title{TinyML for Speech Recognition}

\author{\IEEEauthorblockN{Andrew Barovic\IEEEauthorrefmark{1} and Armin Moin\IEEEauthorrefmark{2}}
\IEEEauthorblockA{\textit{Department of Computer Science, University of Colorado Colorado Springs}\\
United States \\
\IEEEauthorrefmark{1}abarovi2@uccs.edu, \IEEEauthorrefmark{2}amoin@uccs.edu
}
}
    
\maketitle

\begin{abstract}
We train and deploy a quantized 1D convolutional neural network model to conduct speech recognition on a highly resource-constrained IoT edge device. This can be useful in various Internet of Things (IoT) applications, such as smart homes and ambient assisted living for the elderly and people with disabilities, just to name a few examples. In this paper, we first create a new dataset with over one hour of audio data that enables our research and will be useful to future studies in this field. Second, we utilize the technologies provided by Edge Impulse to enhance our model's performance and achieve a high Accuracy of up to 97\% on our dataset. For the validation, we implement our prototype using the Arduino Nano 33 BLE Sense microcontroller board. This microcontroller board is specifically designed for IoT and AI applications, making it an ideal choice for our target use case scenarios. While most existing research focuses on a limited set of keywords, our model can process 23 different keywords, enabling complex commands.
\end{abstract}


\section{Introduction}\label{introduction}
Natural Language Processing (NLP) and Speech Recognition are crucial domains in Artificial Intelligence (AI). While NLP deals with enabling computers to analyze, understand, reason on, and generate human language in textual form, speech recognition is concerned with that in spoken form. It typically transforms the spoken form into a textual modality and processes it. Machine Learning (ML) is a sub-discipline of AI whose methods and techniques are widely deployed in NLP and speech recognition, among other domains. One particular family of ML models is Artificial Neural Networks (ANNs). Deep ANNs possess multiple hidden layers and are much more capable models when trained with large amounts of data.

The training phase typically requires considerable computational resources. Once trained, however, ML models can carry out some tasks, such as making predictions, with a lower expectation of computational resources. Nevertheless, their sizes are still typically too large to fit into the main memory of many resource-constrained Internet of Things (IoT) devices, such as embedded microcontrollers in sensor motes, which possess only a few Kilobytes of memory. With the advent and rapid expansion of the IoT, this has become a more interesting and relevant problem in many domains. 

One may ask why ML models should be deployed on the IoT edge devices as opposed to the much more capable cloud servers. It is reasonable to say that, in most cases, cloud computing is the appropriate solution to ML. However, specific use cases exist in which edge analytics is the more reliable and efficient choice. For instance, you can reduce the network throughput, increase the availability of service, reduce the delays and uncertainties in response time, and, in some cases, ensure data privacy (when the data resides on the user's side rather than in the cloud) if you can enable some ML functionalities on the user's side, such as on edge devices, instead of cloud servers. 

Let's imagine an elderly patient relying on an ML-based prediction service for their healthcare, running on a smartwatch. If the network is down, it would still be desirable to be able to provide some functionality even if there has to be a reduction in the quality of service (e.g., a lower Accuracy). Edge analytics and TinyML can be very beneficial here. Another example is the wake-up command for iOS and Android devices. For instance, to start using the AI-enabled iOS assistant, Siri, the user can typically say \lq{}Hey Siri\rq{}. The speech recognition task to process this wake-up command is not performed in the cloud. It is performed locally on the device to enable fast response and efficient ongoing listening and processing. It is not performed on the main chip, either. The reason is that the main chip should go to sleep/idle or low-power mode whenever the device is not actively in use for energy efficiency reasons. However, a separate \textit{Always On Processor (AOP)}, which is a small, low-power auxiliary processor, continuously listens to the user via the microphone to process any wake-up command. This paper is interested in highly resource-constrained platforms such as the AOP. However, we show that our approach can enable processing far more complex voice commands with 23 different keywords and their combinations.

\textit{TinyML} is a recently emerged sub-field of ML, which involves ML models that can be deployed on microcontrollers with energy consumption in the order of one milli-watt (mW) and main memory in size in the order of Kilobytes. TinyML can enable and support an enormous number of use cases in the IoT \cite{DuttaBharali2021}. The main challenge is to make ML models compact enough to fit into the limited memory and consume a very low amount of energy without compromising the performance (e.g., drop in Accuracy, Precision, and Recall metrics) beyond an acceptable level. Note that compacting ML models often implies applying \textit{quantization} techniques to numeric values (e.g., parameters' weights) that inevitably degrade their performance capabilities. However, our experiments show that if carried out properly, the trade-off is still acceptable enough for many applications.

In this paper, we concentrate on ML models for speech recognition on a resource-constrained IoT platform, namely the \textit{Arduino Nano 33 BLE Sense} \cite{ArduinoNano33BleSense} microcontroller board with a main memory (SRAM) of 256 KB. We use the TensorFlow Lite library \cite{tensorflow-lite, tensorflow-lite-mc}. A state-of-the-art NLP model called MobileBert is included in this library, which has significantly benefited from advanced quantization techniques. Thus, with only a slight drop in Accuracy, its size and speed have become optimized for TinyML applications as a result of applying the quantization techniques \cite{TFliteBlog2020}. Further, we benefit from the services provided by Edge Impulse \cite{edge-impulse}, an ML-Ops online platform that allows data scientists to enhance their ML pipelines for IoT edge devices and TinyML.

The contribution of this paper is twofold: First, we publish an open reference dataset that can be used in future research in this area. This contains an hour of audio data collected using the on-board microphone of the above-mentioned microcontroller board. Second, we propose a novel approach to speech recognition using TinyML that is capable of processing complex commands with 23 keywords. This is far beyond the capabilities of the state of the art in this area.

This paper is structured as follows: Section \ref{related-work} reviews the literature. In Section \ref{proposed-approach}, we propose our novel approach. Section \ref{experimental-results} reports on the experimental results. Moreover, Section \ref{threats-to-validity} points out potential threats to validity. Finally, we conclude and suggest future work in Section \ref{conclusion-future-work}.

\section{Related Work} \label{related-work}
Prior work in the literature, such as Viswanatha et al. \cite{Viswanatha+2022} and Moin et al. \cite{Moin+2022} deployed ML models for speech and hand gesture recognition, as well as for predictive maintenance, respectively, onto an Arduino Nano 33 BLE Sense microcontroller.

Patel et al. \cite{Patel+2023} implemented a self-trained Edge Impulse keyword detection system with a high degree of Accuracy. This model was then deployed on an Arduino Nano 33 BLE Sense microcontroller. 

Waqar et al. \cite{Waqar+2021} used a speech categorization method that could categorize speech and determine the levels of anger in speech, returning a visual reading. 

Nived et al. \cite{Nived+2023} trained a model to recognize custom keywords by using Edge Impulse. They elaborated on the training process and the specific custom keywords selected. Additionally, this research showed a potential method of recognizing more complex keywords provided that the model training was conducted properly. 

Toma et al. \cite{Toma+2020} performed an in-depth exploration of the various structure elements behind the use and function of neural networks in TinyML. Their study covered the tools and processes behind training a model and then and deploying it on a TinyML device. 

Liu et al. \cite{Liu+2024} created a new Deep Neural Network (DNN) model called TinyTS that focused on memory efficiency and speed. 

The Thesis work of Pham \cite{Pham2024} addressed more complex command methods. The implementation in the thesis research was able to recognize 13 total commands and process them with an average accuracy of 80\%. This research is closely related to our work of recognizing more complex speech commands. 

Cioflan et al. \cite{Cioflan+2024} explored a unique method of handling noise featuring on-device ML meant to adapt to background noise, thus increasing Accuracy over time. 

Pimpalkar et al. \cite{PimpalkarNiture2024} studied an application of TinyML that involved person and keyword detection in an elevator use case. 

Raihan Uddin et al. \cite{Raihan-Uddin+2024} conducted a research study on another application of keyword detection involving a smart fan, where the noise, specifically from the fan itself, was a challenge. 

Pavan et al. \cite{Pavan+2024} devised a speaker verification procedure along with keyword detection. 

The thesis of Lin \cite{Lin2024} offered a more efficient approach to TinyML training and inference using some quantization techniques. 

Li et al. \cite{Li+2024} addressed speech recognition and provided test cases and scenarios to test language models. 

Hymel et al. \cite{Hymel+2023} elaborated on the Edge Impulse services and their methods behind the scenes, as well as the Edge Impulse usage in both academia and industry. 

Kiranyaz et al. \cite{Kiranyaz+2021} studied 1D Convolutional Neural Networks (CNNs) as a compact CNN variant useful for our engineering applications of interest in which the ML model compactness is essential. We also adopt this ML model in our work. 

Park et al. \cite{Park+2019} proposed a method of handling speech/audio data. 

Moreover, Theocharides et al. \cite{Theocharides+2024} collected some key insights into the current state of the art regarding TinyML in their conference proceedings summary. 

Kallimani et al. \cite{Kallimani+2024} conducted a comprehensive review of various work related to TinyML.

To our knowledge, none of the existing work in the literature has achieved our level of sophistication for TinyML in speech recognition. We enable a combination of 23 keywords for the speech commands as described in Section \ref{proposed-approach}. We also report our experimental results that outperform the state of the art in Section \ref{experimental-results}.

\section{Proposed Approach} \label{proposed-approach}
We propose a novel approach to speech recognition using ML. We deploy a compact 1D Convolutional Neural Network (CNN) model \cite{Kiranyaz+2021} on a resource-constrained IoT edge platform. The selected platform is an Arduino Nano 33 BLE Sense microcontroller board. This is an affordable choice in terms of price while satisfying the requirements of TinyML in terms of low power consumption and limited memory capacity. 

\begin{figure*}
    \centering
    \includegraphics[width=0.75\textwidth]{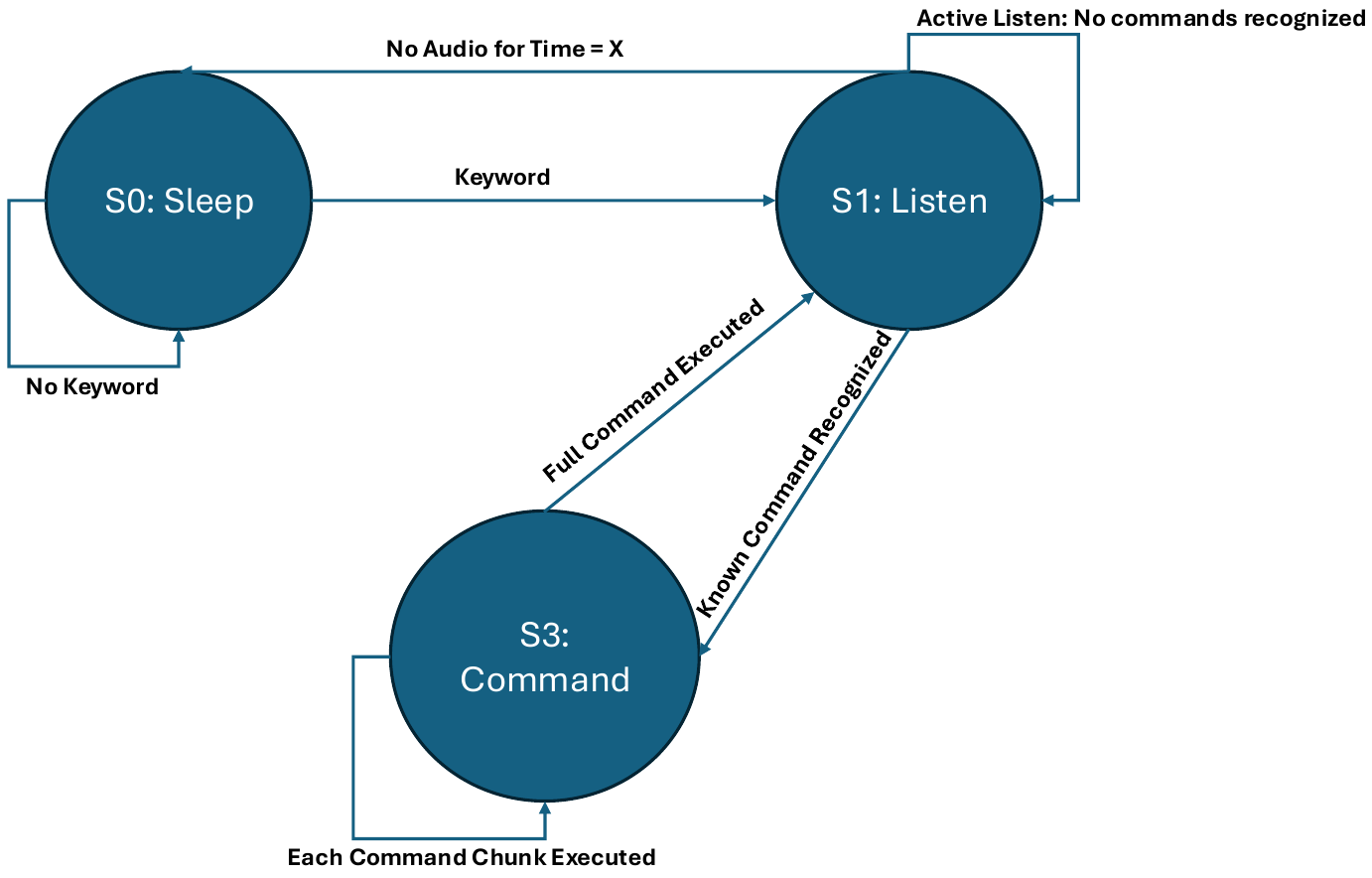}
    \caption{The state diagram showing the behavior architecture of the proposed approach}
    \label{fig:state-logic}
\end{figure*}

The goal is to enable a combination of 23 keywords for voice commands that should control the LEDs on the board. By default, the microcontroller resides in the \textit{sleep} or \textit{idle} mode to save power. In this mode, it keeps listening to the surrounding environment and waiting for a wake-up command that contains the keyword \textit{wake-up}. Once this wake-up command is recognized, the microcontroller switches to the \textit{active} or \textit{listen} mode. In this mode, it tries to recognize specific keywords that make up the actual voice commands to control the LEDs. Whenever no command is received for a certain, predefined period of time, it switches back to sleep or idle mode. Figure \ref{fig:state-logic} illustrates the overall architecture of our approach using a state diagram.

We deploy the ML-Ops online platform \textit{Edge Impulse} \cite{edge_impulse} that is specialized in TinyML for IoT edge devices. We use their services to train efficient and compact ML models and deploy them onto our chip.

\begin{figure*}
    \centering
    \includegraphics[width=0.75\textwidth]{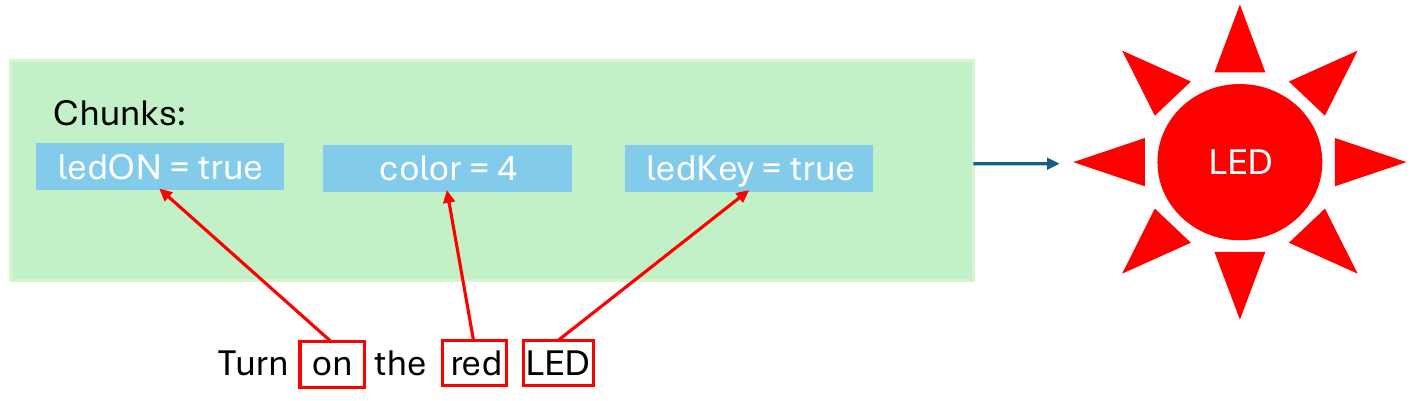}
    \caption{A complex speech command is broken up into smaller chunks, each recognized by one of the two ML models}
    \label{fig:txt-interp}
\end{figure*}

Figure \ref{fig:txt-interp} depicts how a complex speech command is broken up into smaller chunks. Each of these chunks can be processed and, if applicable, recognized by the trained ML model. Each of them sets a flag or variable in the system as a result of the ML model prediction for that chunk.

\section{Experimental Results} \label{experimental-results}
We use an Arduino Nano 33 BLE Sense Rev 2  microcontroller board \cite{ArduinoNano33BleSense}, a micro USB cable capable of data transfer, and a laptop/desktop computer that runs the Arduino 2.3.2 IDE, the Arduino CLI \cite{Arduino_CLI}, Edge Impulse \cite{edge_impulse} in the browser, Edge Impulse CLI, and Node.js. Note that the Arduino IDE needs to have the \textit{Arduino Mbed Nano Boards} library installed. Our software prototype (see \url{https://github.com/qas-lab/BarovicREU}), including a readme file, and our full dataset \cite{dataset} are publicly available.

We begin with a basic hardware test to ensure that the chip is functioning properly. First, we test the Bluetooth functionality of the chip. This is not something needed for this study, but we want to make sure it works for the intended use case scenarios (e.g., smart home and ambient assisted living for the elderly and people with disabilities), which would involve the use of Bluetooth Low Energy (BLE) connectivity to enable efficient communication between the chip and other IoT devices. We also test the LEDs. These initial tests are performed by following a guide provided by Arduino \cite{Bagur}. These tests passed successfully.

Next, we build and train the 1D CNN model using Edge Impulse \cite{edge_impulse, Edge_Impulse_Documentation, Edge_Impulse_CLI}. This ML-Ops platform allows us to easily prepare the data, train the ML model efficiently, and deploy it on the target platform. Additionally, it supports model retraining and testing. One of the techniques that have been deployed for data pre-processing is the Mel-Frequency Cepstral Coefficients (MFCCs) \cite{Tiwari2010}, a typical digital signal processing technique, which improves the ML model's performance.

For each keyword, we have approximately one minute and thirty seconds of recording. We split this into training and testing datasets (80\% training and 20\% testing). Edge Impulse automatically adds its own general noise data for background noise. We used the onboard microphone of the Arduino Nano 33 BLE Sense microcontroller board to collect our dataset.

The dataset is publicly available as an open reference dataset for future use by other studies. This dataset is valuable as it is collected using the actual on-chip microphone rather than a laptop/desktop or other external microphones. This makes the data more similar to the real setup in IoT use cases.

We report the experimental results for speech recognition of the commands using the typical ML performance metrics of Accuracy, Precision, Recall, and F1-Score. Accuracy is defined as the ratio of correctly classified instances to all instances. Our Accuracy was 0.97 on average. Equations \ref{eq:precision}, \ref{eq:recall}, and \ref{eq:f_1} define the Precision, Recall, and F1-Score metrics \cite{Jurafsky_Martin_2024}. Table \ref{tab:eval-results} shows the achieved evaluation results. We only consider ML model predictions that have a confidence of 60\% or higher.

\begin{equation}
Precision = \frac{\text{true positives}}{\text{true positives} + \text{false positives}}
\label{eq:precision}
\end{equation}

\begin{equation}
Recall = \frac{\text{true positives}}{\text{true positives} + \text{false negatives}}
\label{eq:recall}
\end{equation}

\begin{equation}
F_1 = \frac{2 * Precision * Recall}{Precision + Recall}
\label{eq:f_1}
\end{equation}

\begin{table}[ht]
\centering
\caption{Evaluation results (Precision, Recall, and F1-Score). The Accuracy was 0.97 on average.}
\begin{tabular}{|c|c|c|c|c|}
  \hline
  & F1-Score & Precision & Recall \\ \hline
  BLUE & 0.96 & 0.96 &  0.96   \\ \hline
  CYAN & 0.95 & 0.95	 & 0.95   \\ \hline
  GREEN & 1.00 & 1.00	 & 1.00   \\ \hline
  LED & 1.00 & 1.00 & 1.00   \\ \hline
  MAGENTA & 0.98 & 1.00 & 0.95   \\ \hline
  OFF & 0.98 & 1.00 & 0.96   \\ \hline
  ON & 1.00 & 1.00 & 1.00   \\ \hline
  RED & 1.00 & 1.00 & 1.00   \\ \hline
  WAKE UP & 0.98 & 0.96 & 1.00   \\ \hline
  WHITE & 0.95 & 1.00 & 0.91  \\ \hline
  YELLOW & 1.00 & 1.00 & 1.00   \\ \hline
  AND & 0.98 & 0.96 & 1.00   \\ \hline
  BLINK & 1.00 & 1.00 &  1.00  \\ \hline
  CANCEL & 1.00 & 1.00 & 1.00   \\ \hline
  FAST & 1.00	& 1.00 & 1.00   \\ \hline
  FLASH & 1.00 & 1.00 & 1.00   \\ \hline
  NOISE & 0.93 & 0.91 & 0.96   \\ \hline
  NOISE2 & 0.94 & 0.97 & 0.92   \\ \hline
  PLUS & 1.00	& 1.00 & 1.00   \\ \hline
  QUICK & 1.00 & 1.00 & 1.00   \\ \hline
  SLOW & 0.98 & 0.96 &  1.00  \\ \hline
  TOGGLE & 1.00 & 1.00 & 1.00    \\ \hline
  UNKNOWN & 0.98 & 1.00 & 0.96   \\ \hline
\end{tabular}
\\
  \centering
\label{tab:eval-results}
\end{table}

Here is the semantics of the speech command keywords shown in Tables \ref{tab:eval-results} and \ref{tab:train-test-split}:

\begin{itemize}
    \item BLUE: Sets the color variable to 1, indicating blue. The LED will turn blue if this is the last color heard.
    \item GREEN: Sets the color variable to 2, indicating green. The LED will turn green if this is the last color heard.
    \item CYAN: Sets the color variable to 3, indicating cyan. The LED will turn cyan if this is the last color heard.
    \item RED: Sets the color variable to 4, indicating red. The LED will turn red if this is the last color heard.
    \item MAGENTA: Sets the color variable to 5, indicating magenta. The LED will turn magenta/purple if this is the last color heard.
    \item YELLOW: Sets the color variable to 6, indicating yellow. The LED will turn yellow if this is the last color heard.
    \item WHITE: Sets the color variable to 7, indicating white. The LED will turn white if this is the last color heard.
    \item LED: The activation keyword for the LED. It processes the current flags and performs a function. For instance, if color = 1 and ledON = true, the LED will turn blue.
    \item OFF: Sets the ledOFF flag to true and sets color = 0. The LED will turn off if this is the last heard state.
    \item ON: Sets the ledON flag to true. The LED will turn on with the designated color if this is the last heard state.
    \item WAKE UP: Sets the wakeUp flag to true. The device will only start listening when the wakeUp flag is set to true and automatically stops listening after some predefined time.
    \item AND: Sets the andKey flag to true, which allows for combining colors (e.g., red + blue).
    \item CANCEL: Sets the cancelKey flag to true and resets all flags in the code excluding the wakeUp flag.
    \item BLINK: Sets the blinkKey flag to true. This is intended to blink the LED at a specific interval of the color/colors selected.
    \item FAST: Sets the fastKey flag to true to cause quick blinking.
    \item FLASH: Sets the flashKey flag to true to cause a very quick flashing with a specific color.
    \item SLOW: Sets the slowKey flag to true. It is intended to modify the FLASH functionality to blink slow.
    \item NOISE: No specific functionality intended. This is to catch any non-keyword inputs. NOISE is auto-generated by Edge Impulse. This is different from UNKNOWN and NOISE2 (see below).
    \item UNKNOWN: No functionality intended. Similar to NOISE, this is auto-generated by Edge Impulse. However, instead of noise, it contains a random collection of audio files in different languages that have been heavily distorted and some weird sounds.
    \item NOISE2: No specific functionality intended. This label contains the audio that was recorded by us via the onboard microphone, which has an extremely poor quality, and also included a constant environmental noise in the background in our lab environment. Therefore, this is effectively the normal state of the microphone without any speech involved.
    \item PLUS: Sets the plusKey flag to true. It is intended to allow two different commands to function simultaneously.
    \item QUICK: Sets the quickKey flag to true. It is intended to enable a priority mechanism for the commands.
    \item TOGGLE: Sets the toggleKey flag to true. It is intended to toggle the color state of a specific LED. Thus, if an LED is showing, for example, red and green at the same time, and the toggle green command is received, the LED will become red.
\end{itemize}

For the following commands, the speech recognition functionality works but the logic is not implemented in the existing prototype: BLINK, FAST, SLOW, PLUS, QUICK, TOGGLE.

The entire collected dataset contains 1 hour, 3 minutes, and 31 seconds of recording for all samples. Table \ref{tab:train-test-split} illustrates the number of samples for each keyword, as well as the training and testing segregation. Each sample file comprises a one-second interval. Therefore, the number of samples shown in Table \ref{tab:train-test-split} also represents the total recording time in seconds for each keyword. Each keyword has its own train and test split ratio since, for practical reasons, it was not feasible to make them all exactly equal based on the mentioned 80\%-20\% Train-Test split. However, they average out to the target 80\%-20\% split.

\begin{table}[ht]
\centering
\caption{Collected data dedicated to each keyword and the train/test split}
\begin{tabular}{|c|c|c|c|c|}
  \hline
  Keyword & Number of samples & Train/test split (\%) \\ \hline
  BLUE & 116  & 78 / 22    \\ \hline
  CYAN & 104  & 79 / 21   \\ \hline
  GREEN & 104 & 75 / 25   \\ \hline
  LED & 112 & 77 / 23   \\ \hline
  MAGENTA & 104 & 80 / 20  \\ \hline
  OFF & 120 & 77 / 23   \\ \hline
  ON & 129 & 81 / 19   \\ \hline
  RED & 109 & 78 / 22   \\ \hline
  WAKE UP & 112 & 80 / 20   \\ \hline
  WHITE & 120 & 82 / 18  \\ \hline
  YELLOW & 116 & 81 / 19   \\ \hline
  AND & 142 & 81 / 19   \\ \hline
  BLINK & 141 & 85 / 15   \\ \hline
  CANCEL & 134 & 80 / 20   \\ \hline
  FAST & 138 & 81 / 19   \\ \hline
  FLASH & 147 & 81 / 19   \\ \hline
  NOISE & 400 & 80 / 20   \\ \hline
  NOISE2 & 504 & 79 / 21   \\ \hline
  PLUS & 149 &  79 / 21  \\ \hline
  QUICK & 146 & 79 / 21   \\ \hline
  SLOW & 122 & 79 / 21   \\ \hline
  TOGGLE & 142 & 80 / 20   \\ \hline
  UNKNOWN & 377 &  81 / 19  \\ \hline
\end{tabular}
\\
  \centering
  \label{tab:train-test-split}
\end{table}

The achieved experimental results in our study show a high Accuracy, and F-1 Score across all keywords. The lowest F-1 Score being 0.93, was observed in the classification of the noise label that was auto-generated by Edge Impulse. The average F-1 Score was 0.98, the average Precision was 0.97, and the average Recall was 0.98. All in all, we outperform the state of the art. Additionally, our approach can recognize a much larger number of keywords compared to the related work in the literature.

We acknowledge that we decided to use the onboard microphone to make the data similar to the real-world scenario of the intended use cases. However, it is clear that this does not provide a high quality of recording compared to the embedded microphones in typical laptops or external microphones.

\section{Threats to Validity} \label{threats-to-validity}
We collected our dataset using one individual's voice. This poses challenges and limitations in generalization and may induce biases based on sex/gender and other factors. In the future, one may include a more diverse set of recordings for each keyword to mitigate this risk.

Furthermore, the NOISE and UNKNOWN class instances are auto-generated and randomly split by Edge Impulse, leading to a blackbox situation for our method. In general, Edge Impulse is extremely advanced but is not optimal in the transparency of the applied methods. More research is required to understand these aspects of the ML pipeline, as well as the exact quantization techniques.

Also, we have validates this work with one TinyML platform. More research is required to explore other platforms, and a true IoT scenario involving the communication and collaboration between various resource-constrained IoT platforms.

\section{Conclusion and Future Work} \label{conclusion-future-work}

In this paper, we have collected a new dataset and proposed a novel approach to speech recognition on TinyML platforms, highly resource-constrained microcontrollers with energy consumptions in the order of one milliwatt (mW) and main memory capacities in the order of Kilobytes (KB). Our dataset has been useful in this study and will be beneficial to future research work in this area since no other comparable dataset exists for speech recognition on TinyML platforms. Note that we have collected data using the onboard microphone of the TinyML board.

We have trained and deployed a 1D Convolutional Neural Network (CNN) machine learning model to conduct speech recognition. The model can recognize 23 keywords. Moreover, complex commands, such as combinations of two voice commands, are enabled. We have achieved excellent experimental results for validating our approach using the collected dataset in the lab.

In the future, we will enhance our approach. First, we will implement the logic required to fully support the following the BLINK, FAST, SLOW, PLUS, QUICK, and TOGGLE commands. These are currently recognized but the behavioral logic is missing in the prototype to enable the respective actions that must be taken based on these commands. Second, we will work on diversifying our dataset by involving more students and researchers in our lab and in the surrounding academic environment. Further, we will study the underlying methods and techniques used by Edge Impulse and will try to reverse engineer and re-implement part of the prototype to decrease our reliance level on external service providers, such as Edge Impulse. we also plan to work on more advanced concepts, such as transfer learning and federated learning to extend this work.

\section*{Software and Data Availability}
The prototype is available under a permissive open-source license at \url{https://github.com/qas-lab/BarovicREU}. The data used for evaluation are also publicly available at \cite{dataset}.

\section*{Acknowledgment}
This material is based upon work supported by the U.S. National Science Foundation (NSF) under Grant No. 2349452. Any opinions, findings, conclusions, or recommendations expressed in this material are those of the authors and do not necessarily reflect the views of the NSF.

\bibliography{refs}

\begin{thebibliography}{10}
\providecommand{\url}[1]{#1}
\csname url@samestyle\endcsname
\providecommand{\newblock}{\relax}
\providecommand{\bibinfo}[2]{#2}
\providecommand{\BIBentrySTDinterwordspacing}{\spaceskip=0pt\relax}
\providecommand{\BIBentryALTinterwordstretchfactor}{4}
\providecommand{\BIBentryALTinterwordspacing}{\spaceskip=\fontdimen2\font plus
\BIBentryALTinterwordstretchfactor\fontdimen3\font minus \fontdimen4\font\relax}
\providecommand{\BIBforeignlanguage}[2]{{%
\expandafter\ifx\csname l@#1\endcsname\relax
\typeout{** WARNING: IEEEtran.bst: No hyphenation pattern has been}%
\typeout{** loaded for the language `#1'. Using the pattern for}%
\typeout{** the default language instead.}%
\else
\language=\csname l@#1\endcsname
\fi
#2}}
\providecommand{\BIBdecl}{\relax}
\BIBdecl

\bibitem{DuttaBharali2021}
L.~Dutta and S.~Bharali, ``{TinyML meets IoT: A comprehensive survey},'' \emph{Internet of Things}, vol.~16, p. 100461, 2021.

\bibitem{ArduinoNano33BleSense}
``{Arduino Nano 33 BLE Sense},'' \url{https://store.arduino.cc/products/arduino-nano-33-ble-sense}, accessed: 2024-05-29.

\bibitem{tensorflow-lite}
``{TensorFlow Lite},'' \url{https://www.tensorflow.org/lite/guide}, accessed: 2024-05-29.

\bibitem{tensorflow-lite-mc}
``{TensorFlow Lite for Microcontrollers},'' \url{https://www.tensorflow.org/lite/microcontrollers}, accessed: 2024-05-29.

\bibitem{TFliteBlog2020}
{TensorFlow Blog}, ``{What’s new in TensorFlow Lite for NLP},'' \url{https://blog.tensorflow.org/2020/09/whats-new-in-tensorflow-lite-for-nlp.html}, 2020, accessed: 2024-05-29.

\bibitem{edge-impulse}
``{Edge Impulse},'' \url{https://edgeimpulse.com}, accessed: 2024-11-27.

\bibitem{Viswanatha+2022}
\BIBentryALTinterwordspacing
V.~Viswanatha, R.~A.C, R.~Prasanna, P.~C. Kakarla, P.~VivekaSimha, and N.~Mohan, ``Implementation of tiny machine learning models on arduino 33 ble for gesture and speech recognition,'' \emph{ArXiv}, vol. abs/2207.12866, 2022. [Online]. Available: \url{https://api.semanticscholar.org/CorpusID:251067231}
\BIBentrySTDinterwordspacing

\bibitem{Moin+2022}
A.~Moin, M.~Challenger, A.~Badii, and S.~Günnemann, ``Supporting ai engineering on the iot edge through model-driven tinyml,'' in \emph{2022 IEEE 46th Annual Computers, Software, and Applications Conference (COMPSAC)}, 2022, pp. 884--893.

\bibitem{Patel+2023}
\BIBentryALTinterwordspacing
P.~Patel, N.~Gupta, and S.~Gajjar, ``Real time voice recognition system using tiny ml on arduino nano 33 ble *,'' in \emph{2023 IEEE International Symposium on Smart Electronic Systems (iSES)}, 2023, pp. 385--388. [Online]. Available: \url{https://doi.org/10.1109/iSES58672.2023.00085}
\BIBentrySTDinterwordspacing

\bibitem{Waqar+2021}
\BIBentryALTinterwordspacing
D.~M. Waqar, T.~S. Gunawan, M.~A. Morshidi, and M.~Kartiwi, ``Design of a speech anger recognition system on arduino nano 33 ble sense,'' in \emph{2021 IEEE 7th International Conference on Smart Instrumentation, Measurement and Applications (ICSIMA)}, 2021, pp. 64--69. [Online]. Available: \url{http://doi.org/10.1109/ICSIMA50015.2021.9526323}
\BIBentrySTDinterwordspacing

\bibitem{Nived+2023}
\BIBentryALTinterwordspacing
B.~V. Nived, K.~Jamal, G.~Mahesh, and R.~M. Kumar, ``Design of custom keyword recognition using edge impulse on arduino nano 33 ble sense,'' in \emph{2023 2nd International Conference on Applied Artificial Intelligence and Computing (ICAAIC)}, 2023, pp. 1522--1529. [Online]. Available: \url{https://doi.org/10.1109/ICAAIC56838.2023.10140757}
\BIBentrySTDinterwordspacing

\bibitem{Toma+2020}
C.~Toma, M.~Popa, and M.~Doinea, ``Ai neural networks inference into the iot embedded devices using tinyml for pattern detection within a security system,'' in \emph{International Conference on Informatics in Economy Education, Research and Business Technologies}, 2020, pp. 14--22.

\bibitem{Liu+2024}
\BIBentryALTinterwordspacing
Y.-Y. Liu, H.-S. Zheng, Y.~Fang~Hu, C.-F. Hsu, and T.~T. Yeh, ``{TinyTS}: {Memory}-{Efficient} {TinyML} {Model} {Compiler} {Framework} on {Microcontrollers},'' in \emph{2024 {IEEE} {International} {Symposium} on {High}-{Performance} {Computer} {Architecture} ({HPCA})}, Mar. 2024, pp. 848--860, iSSN: 2378-203X. [Online]. Available: \url{https://ieeexplore.ieee.org/abstract/document/10476479}
\BIBentrySTDinterwordspacing

\bibitem{Pham2024}
\BIBentryALTinterwordspacing
D.~A. Pham, ``\BIBforeignlanguage{en}{Implementation of a {Speech}-command-interface on {Microcontroller} with {TinyML}},'' Thesis, Hochschule für Angewandte Wissenschaften Hamburg, Apr. 2024, accepted: 2024-04-05T08:15:22Z. [Online]. Available: \url{https://reposit.haw-hamburg.de/handle/20.500.12738/15406}
\BIBentrySTDinterwordspacing

\bibitem{Cioflan+2024}
\BIBentryALTinterwordspacing
C.~Cioflan, L.~Cavigelli, M.~Rusci, M.~de~Prado, and L.~Benini, ``On-{Device} {Domain} {Learning} for {Keyword} {Spotting} on {Low}-{Power} {Extreme} {Edge} {Embedded} {Systems},'' Mar. 2024, arXiv:2403.10549 [cs, eess]. [Online]. Available: \url{http://arxiv.org/abs/2403.10549}
\BIBentrySTDinterwordspacing

\bibitem{PimpalkarNiture2024}
\BIBentryALTinterwordspacing
A.~S. Pimpalkar and D.~V. Niture, ``Towards {Contactless} {Elevators} with {TinyML} using {CNN}-based {Person} {Detection} and {Keyword} {Spotting},'' May 2024, arXiv:2405.13051 [cs]. [Online]. Available: \url{http://arxiv.org/abs/2405.13051}
\BIBentrySTDinterwordspacing

\bibitem{Raihan-Uddin+2024}
\BIBentryALTinterwordspacing
M.~Raihan~Uddin, A.~Asaduzzaman, K.~Le, and R.~R. Medarametla, ``Voice {Activated} {Edge} {Devices} {Using} {Tiny} {Machine} {Learning} {Enabled} {Microcontroller},'' in \emph{2024 {IEEE} {Green} {Technologies} {Conference} ({GreenTech})}, Apr. 2024, pp. 38--42, iSSN: 2166-5478. [Online]. Available: \url{https://ieeexplore.ieee.org/abstract/document/10520459}
\BIBentrySTDinterwordspacing

\bibitem{Pavan+2024}
\BIBentryALTinterwordspacing
M.~Pavan, G.~Mombelli, F.~Sinacori, and M.~Roveri, ``{TinySV}: {Speaker} {Verification} in {TinyML} with {On}-device {Learning},'' Jun. 2024, arXiv:2406.01655 [cs, eess]. [Online]. Available: \url{http://arxiv.org/abs/2406.01655}
\BIBentrySTDinterwordspacing

\bibitem{Lin2024}
\BIBentryALTinterwordspacing
J.~Lin, ``\BIBforeignlanguage{en}{Efficient {Deep} {Learning} {Computing}: {From} {TinyML} to {LargeLM}},'' Thesis, Massachusetts Institute of Technology, Feb. 2024, accepted: 2024-03-21T19:09:19Z. [Online]. Available: \url{https://dspace.mit.edu/handle/1721.1/153837}
\BIBentrySTDinterwordspacing

\bibitem{Li+2024}
\BIBentryALTinterwordspacing
Y.~Li, J.~Ren, Y.~Wang, G.~Wang, X.~Li, and H.~Liu, ``\BIBforeignlanguage{en}{Audio–visual keyword transformer for unconstrained sentence-level keyword spotting},'' \emph{\BIBforeignlanguage{en}{CAAI Transactions on Intelligence Technology}}, vol.~9, no.~1, pp. 142--152, 2024, \_eprint: https://onlinelibrary.wiley.com/doi/pdf/10.1049/cit2.12212. [Online]. Available: \url{https://onlinelibrary.wiley.com/doi/abs/10.1049/cit2.12212}
\BIBentrySTDinterwordspacing

\bibitem{Hymel+2023}
\BIBentryALTinterwordspacing
S.~Hymel, C.~Banbury, D.~Situnayake, A.~Elium, C.~Ward, M.~Kelcey, M.~Baaijens, M.~Majchrzycki, J.~Plunkett, D.~Tischler, A.~Grande, L.~Moreau, D.~Maslov, A.~Beavis, J.~Jongboom, and V.~J. Reddi, ``Edge {Impulse}: {An} {MLOps} {Platform} for {Tiny} {Machine} {Learning},'' Apr. 2023, arXiv:2212.03332 [cs]. [Online]. Available: \url{http://arxiv.org/abs/2212.03332}
\BIBentrySTDinterwordspacing

\bibitem{Kiranyaz+2021}
\BIBentryALTinterwordspacing
S.~Kiranyaz, O.~Avci, O.~Abdeljaber, T.~Ince, M.~Gabbouj, and D.~J. Inman, ``{1D} convolutional neural networks and applications: {A} survey,'' \emph{Mechanical Systems and Signal Processing}, vol. 151, p. 107398, Apr. 2021. [Online]. Available: \url{https://www.sciencedirect.com/science/article/pii/S0888327020307846}
\BIBentrySTDinterwordspacing

\bibitem{Park+2019}
\BIBentryALTinterwordspacing
D.~S. Park, W.~Chan, Y.~Zhang, C.-C. Chiu, B.~Zoph, E.~D. Cubuk, and Q.~V. Le, ``{SpecAugment}: {A} {Simple} {Data} {Augmentation} {Method} for {Automatic} {Speech} {Recognition},'' in \emph{Interspeech 2019}, Sep. 2019, pp. 2613--2617, arXiv:1904.08779 [cs, eess, stat]. [Online]. Available: \url{http://arxiv.org/abs/1904.08779}
\BIBentrySTDinterwordspacing

\bibitem{Theocharides+2024}
\BIBentryALTinterwordspacing
T.~Theocharides, C.~Frenkel, and L.~Cavigelli, ``Introduction to the {Special} {Issue} on {TinyML},'' \emph{ACM Transactions on Embedded Computing Systems}, vol.~23, no.~3, pp. 40:1--40:5, May 2024. [Online]. Available: \url{https://dl.acm.org/doi/10.1145/3658375}
\BIBentrySTDinterwordspacing

\bibitem{Kallimani+2024}
\BIBentryALTinterwordspacing
R.~Kallimani, K.~Pai, P.~Raghuwanshi, S.~Iyer, and O.~L.~A. López, ``\BIBforeignlanguage{en}{{TinyML}: {Tools}, applications, challenges, and future research directions},'' \emph{\BIBforeignlanguage{en}{Multimedia Tools and Applications}}, vol.~83, no.~10, pp. 29\,015--29\,045, Mar. 2024. [Online]. Available: \url{https://doi.org/10.1007/s11042-023-16740-9}
\BIBentrySTDinterwordspacing

\bibitem{edge_impulse}
\BIBentryALTinterwordspacing
``The leading edge ai platform.'' [Online]. Available: \url{https://edgeimpulse.com/}
\BIBentrySTDinterwordspacing

\bibitem{Arduino_CLI}
\BIBentryALTinterwordspacing
 [Online]. Available: \url{https://arduino.github.io/arduino-cli/1.0/installation/}
\BIBentrySTDinterwordspacing

\bibitem{dataset}
``{Our open reference dataset for speech recognition using TinyML},'' \url{https://figshare.com/s/50807c77f1acc943e4c3}, accessed: 2024-11-27.

\bibitem{Bagur}
\BIBentryALTinterwordspacing
J.~Bagur, ``Connecting nano 33 ble devices over bluetooth®.'' [Online]. Available: \url{https://docs.arduino.cc/tutorials/nano-33-ble-sense/ble-device-to-device/}
\BIBentrySTDinterwordspacing

\bibitem{Edge_Impulse_Documentation}
\BIBentryALTinterwordspacing
 [Online]. Available: \url{https://docs.edgeimpulse.com/docs/edge-ai-hardware/mcu/arduino-nano-33-ble-sense}
\BIBentrySTDinterwordspacing

\bibitem{Edge_Impulse_CLI}
\BIBentryALTinterwordspacing
 [Online]. Available: \url{https://docs.edgeimpulse.com/docs/tools/edge-impulse-cli/cli-installation}
\BIBentrySTDinterwordspacing

\bibitem{Tiwari2010}
V.~T. Tiwari, ``Mfcc and its applications in speaker recognition,'' \emph{Int. J. Emerg. Technol.}, vol.~1, 01 2010.

\bibitem{Jurafsky_Martin_2024}
\BIBentryALTinterwordspacing
D.~Jurafsky and J.~H. Martin, ``Speech and language processing (3rd ed. draft) dan jurafsky and james h. martin,'' Feb 2024. [Online]. Available: \url{https://web.stanford.edu/~jurafsky/slp3/}
\BIBentrySTDinterwordspacing

\end{thebibliography}
\bibliographystyle{IEEEtran}

\end{document}